\documentclass{raa}
\usepackage{graphicx,times}
\usepackage{natbib}
\usepackage{color}
\usepackage{amssymb,amsmath}
\bibpunct{(}{)}{;}{a}{}{,}

\usepackage[a4paper=true,dvipdfm=true,pagebackref=true]{hyperref}
\hypersetup{pdftitle = The title of my PDF, pdfauthor = My name, pdfsubject= The subject, pdfkeywords = keyword1 keyword2 keyword3}
\hypersetup{colorlinks = true, linkcolor = green, anchorcolor = red, citecolor = blue, filecolor = red, pagecolor = red, urlcolor = red}

\begin{document}

   \title{A study of variable stars in the open cluster NGC 1582
   and its surrounding field}

 \volnopage{ {\bf XXXX} Vol.\ {\bf X} No. {\bf XX}, 000--000}
   \setcounter{page}{1}

   \author{Fang-Fang Song\inst{1,2}, Ali Esamdin\inst{1}, Lu Ma\inst{1}, Jin-Zhong Liu\inst{1}, Yu Zhang\inst{1}, Hu-Biao Niu\inst{1}, Tao-Zhi Yang\inst{1,2}
   }

   \institute{ Xinjiang Astronomical Observatory, Chinese Academy of Sciences, Urumqi, Xinjiang 830011, China; {\it aliyi@xao.ac.cn}\\
    \and
    University of Chinese Academy of Sciences, Beijing 10049, China; {\it songfangf@xao.ac.cn}\\
\vs \no
 }

\abstract{ This paper presents the Charge-Coupled Device
time-series photometric observations of the open cluster NGC 1582
and its surrounding field with Johnson $B$, $V$ and $R$ filters
by using Nanshan 1m telescope of Xinjiang Astronomical Observatory. 19 variable stars and 3 variable candidates were
detected in a $45'\times48.75'$ field around the cluster. 12 of the variable stars are newly-discovered variable objects. The physical properties, classifications, and membership of the 22 objects are studied through their light curves, their positions on
the color-magnitude diagram, and with the archival data from
the Naval Observatory Merged Astrometric Dataset.  Among these objects, 5 are eclipsing binary systems, 6 are pulsating variable stars including
one known $\delta$ Scuti star and one newly-discovered RR Lyrae star. The distance of the RR Lyrae star is estimated to be $7.9 \pm 0.3$ kpc, indicates that the star locates far behind the cluster. 4 variable stars are probable members of the cluster, and 13 of the 22 objects are confirmed to be field stars.}

   \authorrunning{F.-F. Song et al. }            %author_head in even pages
   \titlerunning{A study of variable stars in NGC 1582}  % title_head in odd pages
   \maketitle

%________________________________________________ sections below
%
\section{Introduction}           %% first-level sections will be auto-capitalized
\label{sect:intro} Open clusters are the main components of the
stellar population in our Galaxy and play a crucial role in
astrophysical studies (\citealt{Dias+etal+2002}; \citealt{Kharchenko+etal+2005}; \citealt{Piskunov+etal+2006}; \citealt{Kharchenko+etal+2013}). The cluster variables show the
same age, chemical composition, reddening and distance owing to
the common origin they shared (\citealt{Friel+1995}; \citealt{Piskunov+etal+2006}; \citealt{Hasan+etal+2008}).
Searching for variable stars in the open clusters can provide a
window to explore the stellar interiors, verify the stellar
evolution theory, and also offer very important clues for further
understanding of the structure and the evolution of the Milky Way (\citealt{Dias+etal+2002}; \citealt{Piskunov+etal+2006}).

NGC 1582 ($\alpha=04^{\rm h}32^{\rm m}15.4^{\rm s}$, $\delta=43\dg
50'43.0''$; C 0428+437) is a large, sparse open cluster in the Perseus
(\citealt{Baume+etal+2003}). According to the Trumpler system, it
can be classified as type IV2p (\citealt{Trumpler+etal+1930}). The
diameter of NGC 1582 is roughly in the range from $15'$
(\citealt{Dias+etal+2002}) to $37'$ (\citealt{Lynga and
Palous+1987}). It is rarely studied for a small group of bright
stars in the cluster which are mixed with the rich Galactic disk
field star population (\citealt{Carraro+2002}). The basic parameters of NGC 1582 are listed in Table 1.

The first announcement of variables of NGC 1582 was contributed by \citet{Richter+1970}, who discovered two variable stars named as NSV 1620 and NSV 1633. Both of the stars were described with very approximate coordinates, and no existing light curves can be found in the literature. Richter classified NSV 1620 as a star of unknown type, and he believed that NSV 1633 could be an eclipsing variable. All the basic parameters of the stars can be found in the General Catalogue of variable Stars (GCVS\footnote{the General Catalogue of variable Stars,
http://www.sai.msu.su/gcvs/}) (\citealt{Samus+etal+2005}). The first modern
study of NGC 1582 was published by \citet{Baume+etal+2003}. From
the $UBV(I)c$ photometry and the high-resolution spectroscopy for
the bright stars in this cluster, \citet{Baume+etal+2003}
estimated the reddening of $E (B-V) = 0.35 \pm 0.03$ mag, the
approximate distance away from the Sun of $1100 \pm 100$ pc and the
age of $300\pm100$ Myr (\citealt{Baume+etal+2003}). Recently, R.Furgoni reported a $\delta$ Scuti star of NGC 1582 named as VSX J043245.8+434930 with a period of 0.09778 day in 2011 (\citealt{Furgoni+2011}). This star's parameters including its light curve can be found in the VSX database\footnote{The International Variable Star Index, http://www.aavso.org/vsx/}. Then, a Charge-Coupled Device (CCD)  observations of NGC 1582 were conducted by \citet{Yang+etal+2013}. They presented two days of CCD time-series in $V$ and $R$ bands, and announced the discovery of 6 variables in the region of the cluster (\citealt{Yang+etal+2013}). The basic parameters and the light curves of those variables are presented in Table 3 and Fig.8 of \citet{Yang+etal+2013}, however, their variable types were not discussed.

In order to study variable stars in the open cluster NGC 1582 and
its surrounding field, we obtained CCD time-series photometry in
Johnson $B$, $V$ and $R$ bands using Nanshan 1m telescope of
Xinjiang Astronomical Observatory (XAO).

The paper is organized as follows. The
observations are described in Section 2. The data reduction and calibration are given in section 3. The analysis and results are presented in Section 4. We give our conclusions in Section 5.

\begin{table}
\bc
\begin{minipage}[]{100mm}
\caption[]{The basic parameters of NGC
1582\label{tab1}}
\end{minipage} \setlength{\tabcolsep}{1pt} \small
 \begin{tabular}{ccccccccccccc}
  \hline\noalign{\smallskip}
Name & RA & DEC & $l$ &$b$ &Diameter &Distance &Age &Subclass \\
&(2000)&(2000)&(deg)&(deg)&(arcmin)&(pc)&(Myr)&\\
  \hline\noalign{\smallskip}
NGC 1582 & 04:32:15.4 & +43:50:43 & 159.30 & -2.89 & $15 \sim 37$ & $1100 \pm 100$ & $300 \pm 100$ & IV2p \\
  \noalign{\smallskip}\hline
\end{tabular}
\ec
\label{tab1}
\end{table}

\begin{figure}[htbp]
   \centering
   \includegraphics[width=0.9\textwidth, angle=0]{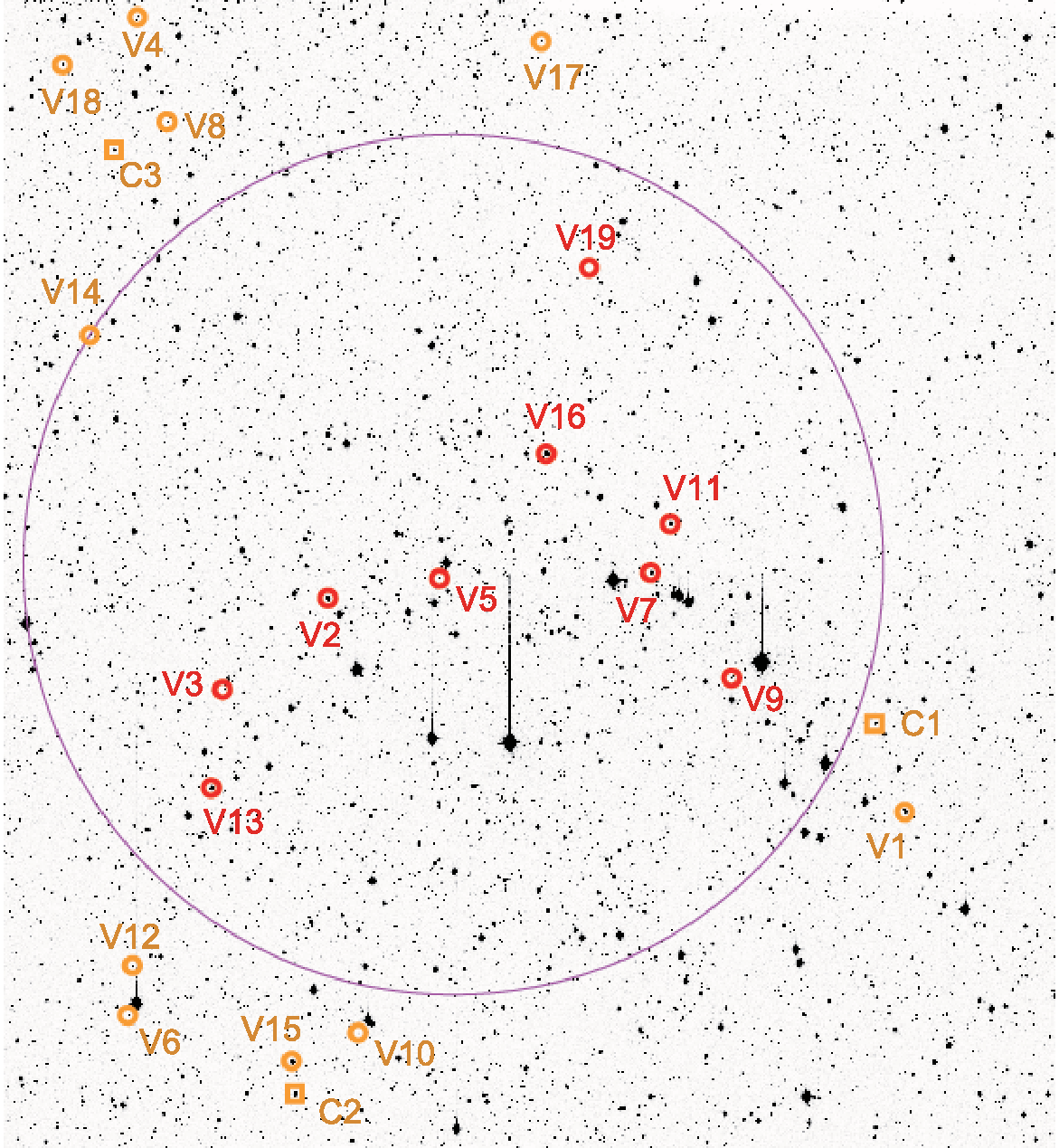}
   \caption{Observation image of NGC 1582 region ($45'\times48.75'$ in R filter). The left is east and the top is north. The
center of the magenta circle is $\alpha=04^{\rm h}32^{\rm m}15.4^{\rm s}$,
$\delta=43\dg 50'43''$.0 (\citealt{Dias+etal+2002}), and the radius of it is $18.5'$ (\citealt{Lynga and
Palous+1987}). The 19 variable stars and 3 variable candidates detected by this work are marked in the image. Nine variables inside the magenta circle are marked in red color, the others are in orange. The red ones are more likely members of the cluster than the orange ones.}
%\end{minipage}
   \label{Fig1}
   \end{figure}

\section{Observations}
\label{sect:Obs}

NGC 1582 was observed through the $B$, $V$ and $R$ filters of the
standard Johnson-Cousin-Bessel multi-color filter system with the
Nanshan 1m telescope of XAO. The telescope sits on an alt-azimuth
mount, and it is primary focus design. While the CCD camera (E2V
CCD203-82 (blue) chip) has $4096\times4096$ pixels, only about a
quarter of the CCD chip with $2400\times2600$ or $2400\times2900$
pixels was used, corresponding to a field of view of
$45'\times48.75'$ or $45'\times54.375'$, respectively. The observed field of NGC 1582 is shown in Fig. 1.

Observations were carried out for 14 nights (61.9 total monitoring
hours) from 2014 December to 2015 January. The exposure times of
the $B$, $V$ and $R$ filters were 30s, 15s, 12s, respectively.
Totally, we collected 1792 CCD frames of NGC 1582, including 599
$B$-band, 594 $V$-band and 599 $R$-band images. The average seeing was 2.5 arc seconds during the observations, and the airmass ranged from 1.1 to 1.6. The journal of our observations is listed in Table 2.

\begin{table}
\bc
\begin{minipage}[]{140mm}
\caption[]{Journal of observations of the open cluster NGC 1582 using
Nanshan 1m telescope.\label{tab2}}\end{minipage}
\setlength{\tabcolsep}{1pt} \small
 \begin{tabular}{ccccccccccccc}
  \hline\noalign{\smallskip}
Date     &     Start    &Length&     Frames     &   Exposure(s) & Seeing &Airmass&    CCD-chip     &\\
         &(HJD 2456900+)& (hr) &($B$, $V$, $R$) &($B$, $V$, $R$)&(arcsec)&       &                 &\\
  \hline\noalign{\smallskip}
2014Dec02&    94.071    & 0.7  & 10,   9 ,   9  & 40,  25,  20  &  2.6   &1.3683 & $2400\times2600$&\\
2014Dec03&    95.036    & 0.4  & 7 ,   7 ,   8  & 30,  15,  12  &  3.1   &1.6349 & $2400\times2600$&\\
2014Dec06&    98.049    & 2.6  & 43,   44,   44 & 30,  15,  12  &  2.3   &1.2608 & $2400\times2600$&\\
2014Dec09&   101.030    & 0.5  & 6 ,   6 ,   6  & 30,  15,  12  &  3.5   &1.5190 & $2400\times2600$&\\
2014Dec11&   103.038    & 1.0  & 15,   16,   16 & 30,  15,  12  &  2.8   &1.3807 & $2400\times2600$&\\
2015Jan01&   124.036    & 9.3  & 60,   59,   59 & 30,  15,  12  &  2.5   &1.2845 & $2400\times2600$&\\
2015Jan02&   125.031    & 1.7  & 24,   23,   23 & 30,  15,  12  &  2.5   &1.1225 & $2400\times2600$&\\
2015Jan04&   127.036    & 8.8  & 88,   88,   83 & 30,  15,  12  &  2.5   &1.2560 & $2400\times2900$&\\
2015Jan05&   128.036    & 6.8  & 66,   66,   66 & 30,  15,  12  &  2.6   &1.1080 & $2400\times2900$&\\
2015Jan06&   129.029    & 8.8  & 85,   85,   84 & 30,  15,  12  &  2.5   &1.2533 & $2400\times2900$&\\
2015Jan07&   130.299    & 2.7  & 36,   37,   35 & 30,  15,  12  &  2.8   &1.6198 & $2400\times2900$&\\
2015Jan08&   131.031    & 9.2  & 76,   76,   75 & 30,  15,  12  &  2.5   &1.3543 & $2400\times2900$&\\
2015Jan09&   132.031    & 9.2  & 83,   82,   83 & 30,  15,  12  &  2.8   &1.3634 & $2400\times2900$&\\
2015Jan10&   133.292    & 0.2  & 4 ,   3 ,   4  & 30,  15,  12  &  3.5   &1.2774 & $2400\times2900$&\\
  \noalign{\smallskip}\hline
\end{tabular}
\ec
%% place \tablecomments and \tablerefs below \end{center| and \end{center}:
%% you may leave the table-width parameter to editors or set to your actual size
\end{table}

\section{Data Reduction and Calibration}

The CCD time-series frames were pre-processed by the
IRAF\footnote{Image Reduction and Analysis Facility,
http://iraf.noao.edu/} routines for subtracting bias,
overscan and applying the flat-field correction. The dark
correction was not considered since the CCD camera was operated at
about $-120 \dg$$C$ with liquid nitrogen cooling and thus the
thermionic noise was less than 1 e $pix^{-1}h^{-1}$ at the
temperature.  The pixel coordinates were converted to the equatorial coordinates by matching with the third US Naval Observatory CCD Astrograph Catalog (UCAC3).

The Nanshan 1m telescope has a wide field of $1.3\dg\times1.3\dg$, the focal plane of the telescope is not flat, leading to the star image of the full field of view would be changed slightly, more obvious at the edge of the pictures. The photometric software Sextractor (\citealt{Bertin+etal+1996}) was therefore employed to perform the aperture photometry. Sextractor can provide the best apertures based on the shapes of star images automatically to ensure the accuracy of photometry of the whole field. The detail information of Sextractor can be consulted from http://www.astromatic.net/ or \citealt{Bertin+etal+1996}. Fig. 2 shows the photometric errors and their trends against $B$, $V$ and $R$ magnitudes of our observations.

\begin{figure}[htbp]
\centering
   \includegraphics[width=0.9\textwidth]{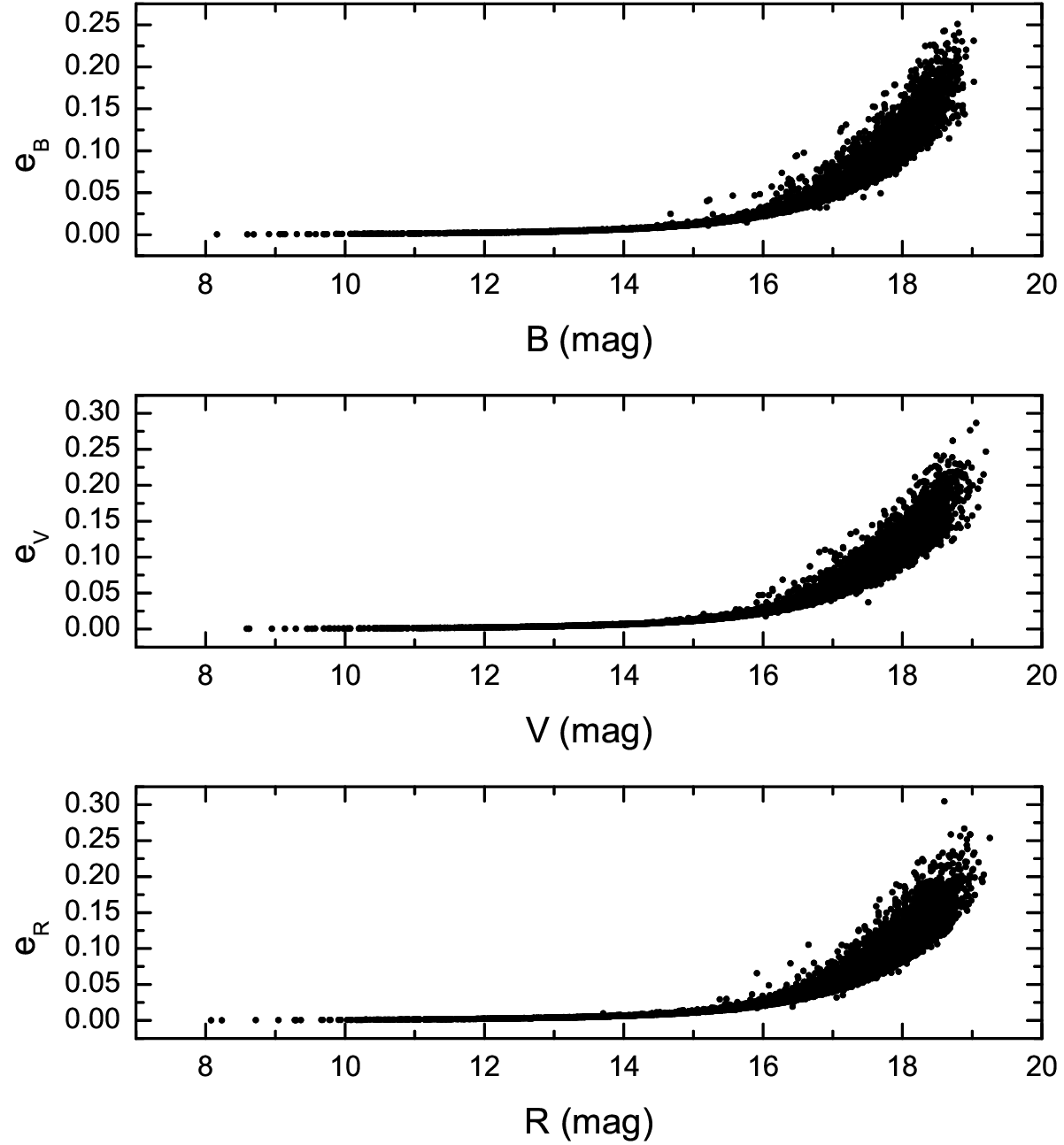}
   \caption{Photometric errors of $B$, $V$, and $R$ bands of the observations.}
   \label{Fig2}
   \end{figure}

Then a data processing system of XAO time-domain survey was used
to acquire the light curves of all stars in the observed field. The data processing system is based on
\citealt{Tamuz+etal+2005}, \citealt{Cameron+etal+2006} and \citealt{Ofir+etal+2010}. The roughly
de-correlated differential magnitude of one star is given by
\begin{equation}
  x_{ij} = m_{ij} - \hat m_{j} - \hat z_{i}
\end{equation}
where $m_{ij}$ is a two-dimensional array of instrumental magnitudes, $i$ denotes a single CCD frame within the entire season¡¯s data, $j$ labels an
individual star. $\hat m_{j}$ is the mean instrumental magnitude for each star, and $\hat z_{i}$ is the zero-point correction for each frame
(\citealt{Cameron+etal+2006}).

In order to search for variables in the observed field, we examined visually all three-band light curves of our collection obtained by above process. The light curves which show significant simultaneous variations and relatively large Root Mean Square (RMS) in all filters are considered as variable stars. Then, we checked all images again to ensure that the possible variables were not located at the frame's edge nor contaminated from bright stars. The RMS vs magnitude diagram of all stars in the $R$ band is shown in Fig. 3.
\begin{figure}[htbp]
   \centering
   \includegraphics[width=0.9\textwidth]{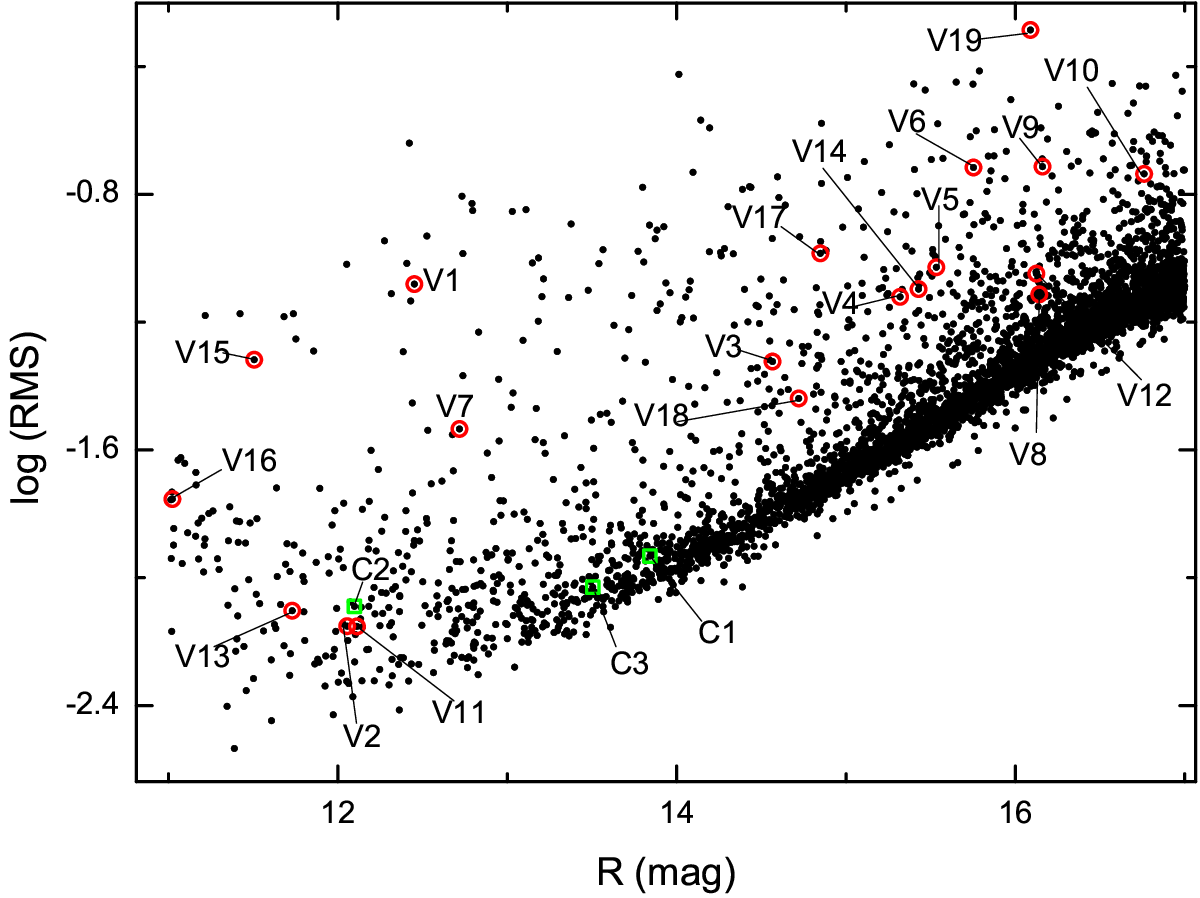}
   \caption{The RMS - magnitude diagram of all light curves obtained in
 $R$ band. The red open circles indicate the 19 variable
stars detected in this work, while the green boxes indicate the 3 variable candidates.}
   \label{Fig3}
   \end{figure}

110 constant stars with photometry precisions better than 0.03 mag
in the $R$-band were selected to do the magnitude calibration. It can
produce comparison stars through the star brightness with random
scattered in the time series. We excluded the stars near to the CCD edges or have been contaminated by other stars or
bad pixels. The instrumental magnitudes $b$, $v$, $r$ were then
converted to the standard system by the least-square fit to straight
lines, whereas the standard magnitudes $B$, $V$, and $R$ were obtained
through a cross match with data of the Naval Observatory
Merged Astrometric Dataset (NOMAD1)
(\citealt{Zacharias+etal+2005}). The calibration formulas for NGC
1582 are:
\begin{equation}
  V-v = 0.19( \pm 0.08)(v - b) + 0.42( \pm 0.02)
\end{equation}
\begin{equation}
  B-b = 0.47( \pm 0.11)(v - b) + 1.16( \pm 0.03)
\end{equation}
\begin{equation}
  R-r = -0.18( \pm 0.11)(v-r) + 0.64( \pm 0.04)
\end{equation}

\section{Results and dicussion}

\subsection{Light curves and observed variable stars}

A total of 19 variable stars, labeled V1, V2, $\cdots$, V19 in this paper, and three candidates C1, C2, C3, have been detected through the process described in Section 3. Out of these 12 are new detections. The location of these 22 stars in the field of the cluster is shown in Fig. 1.

Fig. 3 displays the RMS of detected stars in the field of the cluster and the 22 variables and candidates are indicated with red open circles (19 variables) and green open boxes (3 candidates).

V1 is listed as a variable star in the GCVS named as NSV 1620, and V2 is listed in the VSX database named as VSX J043245.8+434930. After comparing to the catalog of \citet{Yang+etal+2013}, five variables, labeled as V3, V4, V5, V6, V7 in this paper,  were originally discovered by \citet{Yang+etal+2013} and labeled as V1, V2, V4, V5, and V6 in their paper, are confirmed in our observations. The light curves of V3 of \citet{Yang+etal+2013} (V3y for distinction thereafter) and NSV 1633 obtained in this work are shown in Fig. 4. The upper panels show the light curves of V3y and lower panels that of NSV 1633. These light curves do not present variations above 0.07 and 0.05 mag levels respectively, therefore we do not confirm their variable nature. Thus, the 12 variables v8, $\cdots$, V19 are all new discoveries as it is the detection of the three candidates C1, C2 and C3.

\begin{figure}[htbp]
\centering
 \includegraphics[width=1\textwidth]{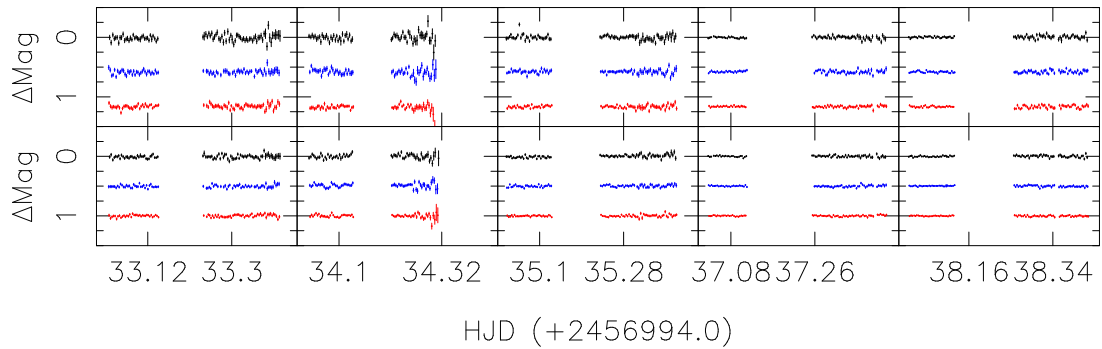}
\caption{Light curves of V3y (upper panels) and NSV 1633 (lower panels) after differential photometry with compare stars in NGC 1582 in the $B$ (black curves), $V$ (blue curves) and $R$ (red curves) filters observed by Nanshan 1 m telescope in the best 5 observation nights.}
 \label{Fig4}
\end{figure}

\begin{figure}[htbp]
\centering
 \includegraphics[width=1\textwidth]{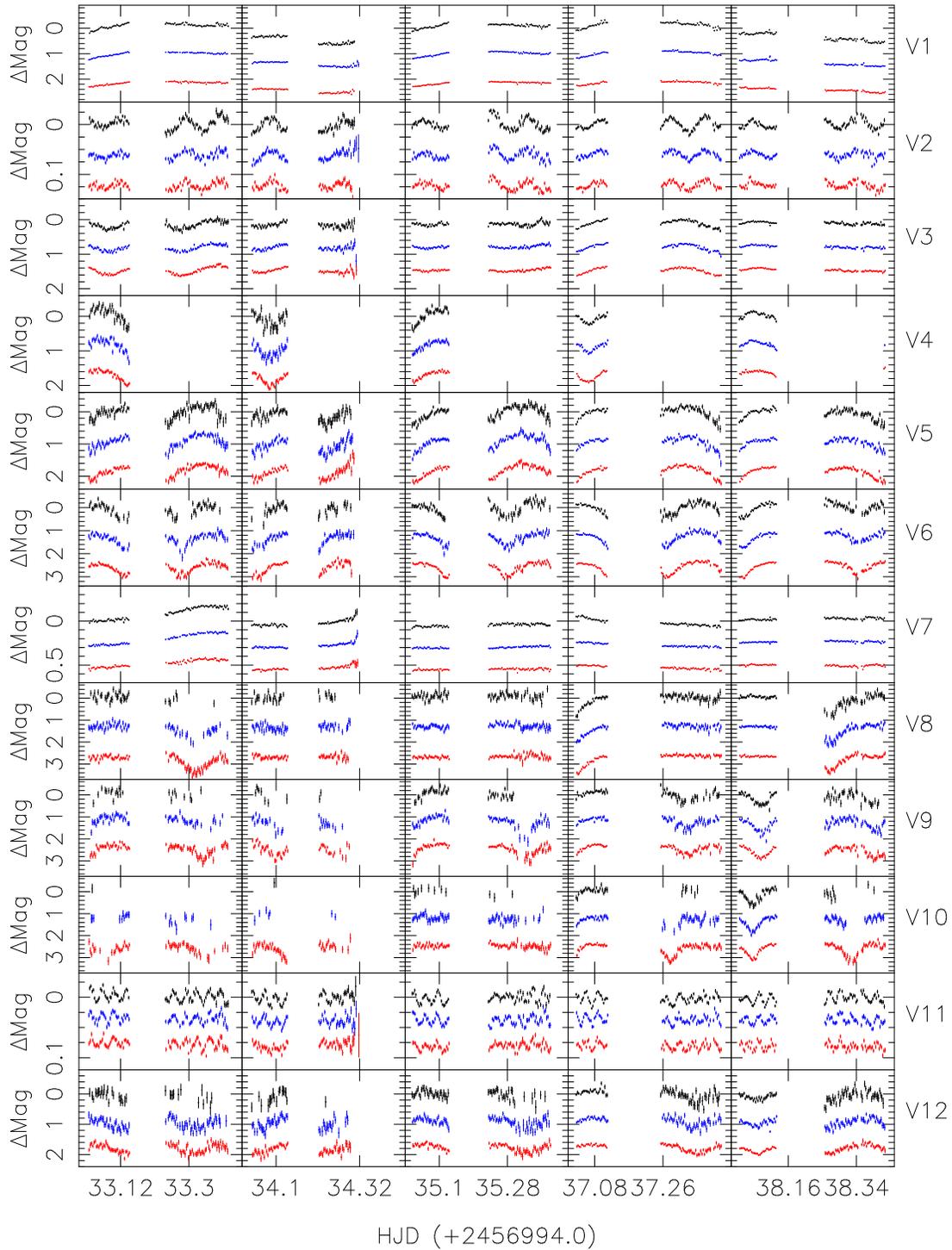}
\caption{Light curves of variable stars and variable candidates after differential photometry
with compare stars in NGC 1582 in the $B$ (black curves), $V$
(blue curves) and $R$ (red curves) filters observed by Nanshan 1m
telescope in the best 5 observation nights.}
\end{figure}
\begin{figure}
\centering
 \includegraphics[width=1\textwidth]{plots-msRAA-2016-0058/msRAA-2016-0058fig5-2.ps}
\textbf(Fig. 5) Continue.
 \label{Fig5}
\end{figure}

The light curves of the 22 objects are shown in Fig. 5. The probable types of variable stars can be specified roughly considering the features of their light curves (\citealt{Samus+etal+2005}; \citealt{Sterken+etal+2005}). As shown in Table 3, We  identify 5 eclipsing binary systems, including 2 Algol (Beta Persei)-type eclipsing systems (marked as ``EA" ) and 3 W Ursae Majoris-type eclipsing systems (marked as ``EW" ). The periods of the eclipsing systems are extracted by the PDM program of IRAF-ASTUTIL, which is based on the phase dispersion minimization algorithm (\citealt{Stellingwerf+1978}). The folded light curves of the 5 eclipsing binary systems are presented in Fig. 6, and the parameters of them are given in Table 3.

\begin{figure}[htbp]
\centering
   \includegraphics[angle=-90,width=0.8\textwidth]{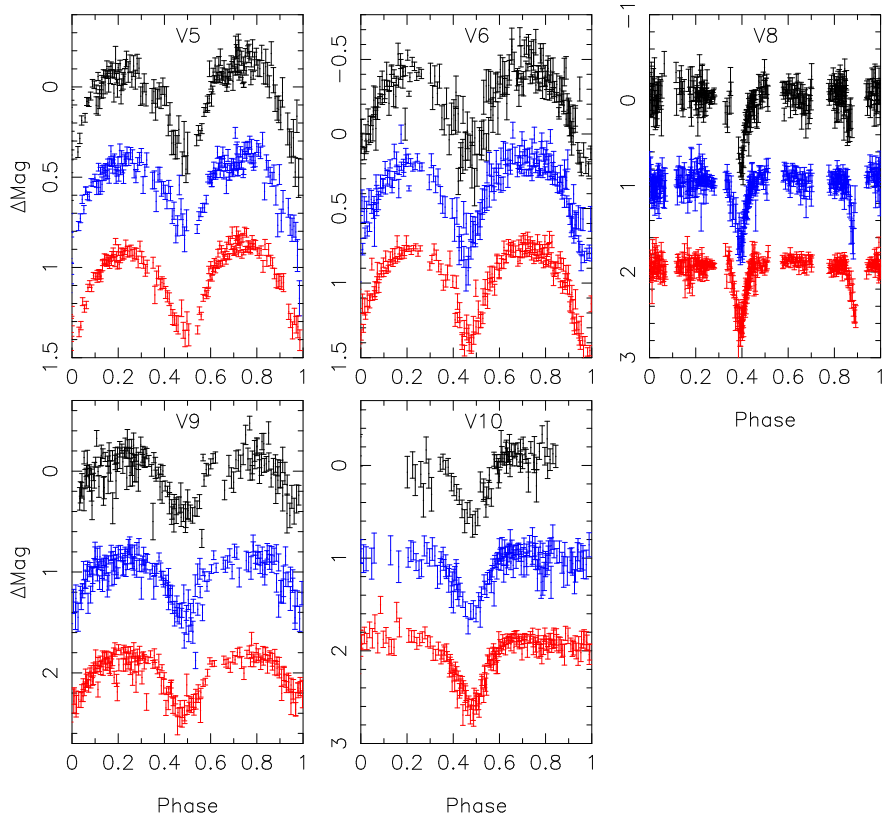}
   \caption{Phased light curves of 5 eclipsing binary systems in the $B$
(black curves), $V$ (blue curves) and $R$ (red curves) bands.}
   \label{Fig6}
   \end{figure}

\begin{table}
\centering
\begin{minipage}[]{120mm}
\caption[]{Parameters of the eclipsing binary systems in the observed
field.}\end{minipage} \setlength{\tabcolsep}{2.5pt}
\small
 \begin{tabular}{ccccccccccccc}
  \hline\noalign{\smallskip}
ID & RA & DEC & Magnitude NOMAD1&  Amp &  P  & Type  & CM$^1$ &Ref$^2$  \\
&(2000) &(2000) & ($B$ ,\, $V$ ,\, $R$) & (mag) &(days)  & & &\\

  \hline\noalign{\smallskip}
V5 &04:32:18.871 &+43:50:09.10&16.927, 16.162, 16.086 &  0.62 &0.3979    &EW    & pm    &V4$^\star$\\
V6 &04:33:37.291 &+43:31:55.81&17.203, 16.410, 16.319 &  0.82 &0.3067    &EW    & nm    &V5$^\star$\\
V8&04:33:19.440 &+44:10:16.39&17.605, 16.797, 16.677 &  1.00 &1.2368     &EA    & nm    &\\
V9&04:31:10.344 &+43:45:18.47&17.830, 16.961, 16.848 &  0.82 &0.3067    &EW    & pm    &\\
V10&04:32:42.833 &+43:30:46.04&18.537, 17.611, 17.352 &  0.92 &0.2624     &EA    & nm   &\\

  \noalign{\smallskip}\hline
\end{tabular}
\flushleft
  $^1$  Cluster membership: `pm' - probable member, `nm' - non-member.   \\
  $^2$ Names given in the discovering paper of \citet{Yang+etal+2013}.
\label{Table 3}
\end{table}

6 variable stars, which are listed in Table 4, show typical characters of pulsating. In order to estimate the periods of these 6 pulsating variables, the phase-match technique and frequency analysis are used. The frequency analysis was performed by PERIOD04 (\citealt{Lenz and Breger+2005}), and the light curves are fitted by
\begin{equation}
  m = m_{0} - \sum A_{i} - sin(2\pi(f_{i}t + \phi_{i}))
\end{equation}
The folded light curves of these pulsating variable stars are shown in Fig. 7. V2 is marked ``DSCT" in Table 4 for it is identified as a $\delta$ Scuti star in the VSX database. Considering the the shape of their light curves and the length of the detected period together with the amplitude, V3, V12 and V13 probably belong to the $\delta$ Scuti class, while V3 and V12 are possible High Amplitude Delta Scuti stars (HADS). V11 is possible a $\delta$ Scuti star or a GW Vir star. The light curves of V14 agrees well with the RR Lyrae RRab, and is marked as ``RR Lyrae" in Table 4.

By adopting $E(B-V) = 0.35$ mag (\citealt{Baume+etal+2003}) and the absolute magnitude of RRab $M_{V} = 0.575\pm0.082$ mag (\citealt{Arellano Ferro+etal+2016}), thus $(m-M)_{0} = 14.5\pm0.09$, then the distance to V14 is estimated to be $7.9 \pm 0.3$ kpc.  The quoted uncertainty is estimated through uncertainties of measured magnitude£¬reddening and absolute magnitude. Since the distance of the cluster is $1.1 \pm 0.1$ kpc (\citealt{Baume+etal+2003}), we can confirm that V14 is a field RR Lyrae behind the cluster.

\begin{figure}[htbp]
\centering
  \includegraphics[angle=-90,width=0.8\textwidth]{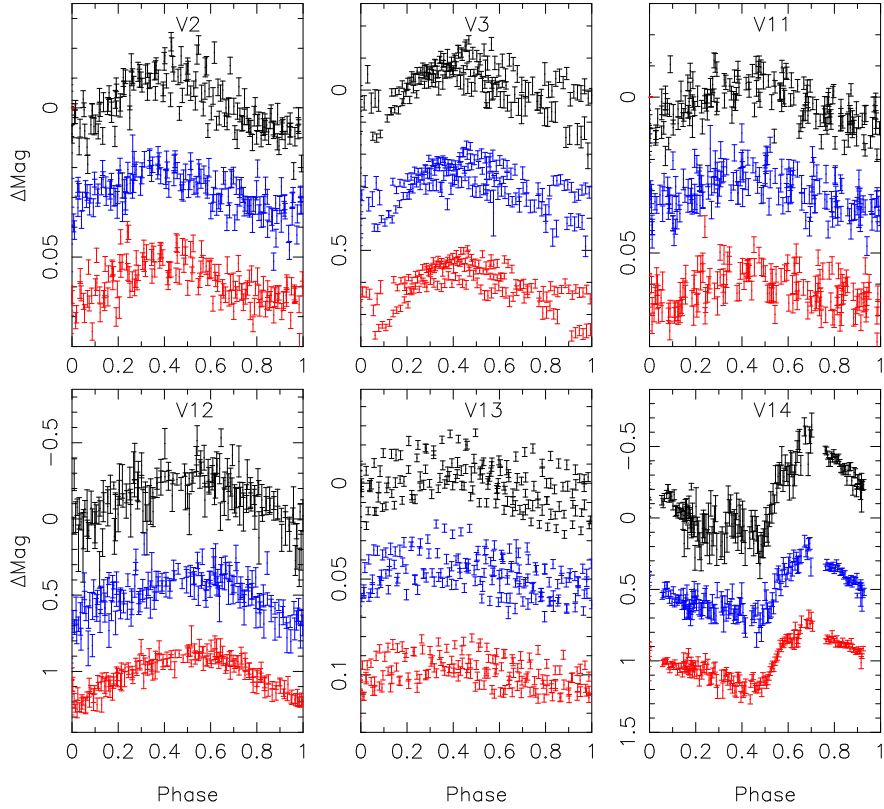}
  \caption{Phased light curves of the 6 pulsating variable stars in the $B$
(black curves), $V$ (blue curves) and $R$ (red curves) bands.}
   \label{Fig7}
\end{figure}

\begin{table}
\bc
\begin{minipage}[]{120mm}
\caption[]{Parameters of the pulsating variable stars in the observed
field.}\end{minipage} \setlength{\tabcolsep}{2.5pt}
\small
 \begin{tabular}{ccccccccccccc}
  \hline\noalign{\smallskip}
ID & RA & DEC & Magnitude NOMAD1&  Amp &  P  & Type  & CM$^1$ &Ref$^2$  \\
&(2000) &(2000) & ($B$ ,\, $V$ ,\, $R$) & (mag) &(days)  & & &\\
  \hline\noalign{\smallskip}
V2 &04:32:45.898 &+43:49:30.72&13.156, 12.570, 12.671&  0.04 &0.0883  &DSCT  &    &$VSX J043245.8+434930^\sharp$  \\
V3 &04:33:11.292 &+43:45:45.79&15.911, 15.158, 15.113&  0.30 &0.1968  &     & pm   &$V1^\star$  \\
V11 &04:31:23.242 &+43:52:03.29&12.979, 12.525, 12.758&  0.02 &0.0312  &    & pm   &  \\
V12&04:33:35.726 &+43:34:03.36&17.719, 16.818, 16.723&  0.56 &0.1833  &     & nm   &  \\
V13&04:33:15.329 &+43:41:33.86&12.856, 12.253, 12.343&  0.04 &0.0838  &     &    &  \\
V14&04:33:39.936 &+44:01:13.80&17.074, 16.160, 15.979&  0.68 &0.5858  &RR Lyrae     & nm   &  \\

  \noalign{\smallskip}\hline
\end{tabular}
\ec
\flushleft
  $^1$  Cluster membership: `pm' - probable member, `nm' - non-member.   \\
  $^2$ Name given in the discovering paper of \citet{Yang+etal+2013} (marked with $\star$) and name given by VSX database (marked with $\sharp$).
\label{Table 4}
\end{table}

The variables presented in Table 5 are the stars whose types
remain unconfirmed. Among these stars, V17 and V18 are likely Algol
(Beta Persei)-type eclipsing systems (marked by ``EA"), since each
one shows one possible eclipse in their light curves. It might be possible that we have missed one or more eclipses while not observing. So, it is impossible for us to estimate a lower limit for the orbital period of the two variables and obtain their phased light curves. V16 is marked as ``E" type since its light curves are similar to that of an eclipsing system. The periods of V1, V7 and V15 can't be determined from the current work, because their periods are most likely longer than 2 days.

The first panel of Fig. 8 gives the phased light curves of V4, and shows that the shape of light curves of V4 is not very similar to a eclipsing binary system, since the different depths of two minima are not visible. The light curves remind the curve for a RRc, but the period is too short for a RR Lyrae star. Thus it could be a monoperiodic high amplitude $\delta$ Scuti star.

The variation of light curves of V19 is quite like a pulsating star, but we could not find a reliable period for it. Obviously, more data and a longer time-series observations are required for this star.

\begin{figure}[htbp]
\centering
  \includegraphics[angle=-90,width=0.8\textwidth]{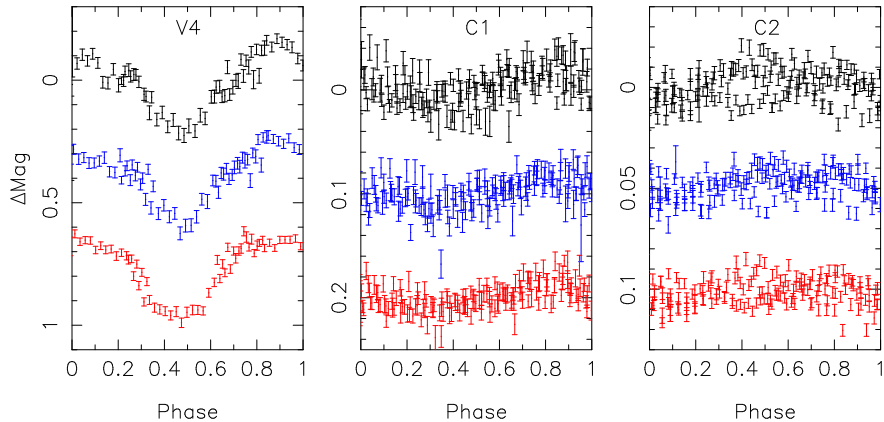}
  \caption{Phased light curves for V4, C1, and C2 in the $B$
(black curves), $V$ (blue curves) and $R$ (red curves) bands.}
   \label{Fig8}
\end{figure}

\begin{table}
\bc
\begin{minipage}[]{120mm}
\caption[]{parameters of the type unconfirmed variable stars in
the observed field.}\end{minipage}
\setlength{\tabcolsep}{2.5pt} \small
 \begin{tabular}{ccccccccccccc}
  \hline\noalign{\smallskip}
ID & RA & DEC & Magnitude NOMAD1&  Amp &  P  & Type  & CM$^1$ &Ref$^2$  \\
&(2000) &(2000) & ($B$ ,\, $V$ ,\, $R$) & (mag) &(days)  & & &\\
  \hline\noalign{\smallskip}
V1 &04:30:30.689 &+43:39:11.38&14.597, 13.506, 13.165&  0.64 &$>$2.00      &     & nm   &$NSV 1620^\sharp$ \\
V4 &04:33:25.531 &+44:14:49.13&16.756, 15.950, 15.872&  0.60 &0.1354   &     & nm   &$V2^\star$ \\
V7 &04:31:28.481 &+43:49:59.88&13.871, 13.257, 13.329&  0.22 &$>$2.17   &     &    &$V6^\star$ \\
V15 &04:32:58.877 &+43:29:40.63&12.984, 12.203, 12.123&  0.20 &$>$2.00      &     & nm   & \\
V16 &04:31:52.121 &+43:55:18.80&11.698, 11.350, 11.687&  0.24 &$>$10.00  &E          &    &  \\
V17&04:31:48.859 &+44:13:02.24&16.178, 15.422, 15.398&  1.16 &      &EA          & nm   &  \\
V18&04:33:43.932 &+44:12:54.22&16.049, 15.317, 15.290&  0.52 &   &EA          & nm   & \\
V19&04:31:39.821 &+44:03:13.64&17.782, 17.018, 17.019&  0.72 &   &     &    & \\

\noalign{\smallskip}\hline
\end{tabular}
\ec
\flushleft
  $^1$  Cluster membership: `nm' - non-member.   \\
  $^2$ Names given in the discovering paper of \citet{Yang+etal+2013} (marked with $\star$) and name given by GCVS (marked with $\sharp$).
\label{Table 5}
\end{table}

Three variable candidates detected in this work are listed in Table 6.  As shown in Fig. 5, variations of C1, C2, and C3 are very marginal. The light curves of C1 and C2 are a bit like those of pulsating stars, but their phased light curves are not obvious (see Fig. 8). C3 is likely a variable star, since it shows a 0.1 mag increase near HJD 2457031.11 in its light curves (see the fourth panel of the bottom panels in Fig. 5). We have checked its images and position and found no evidence to indicate that it is affected by bad conditions. The signal-to-noise ratio of this variation is about 5.9, C3 is thereby classified as a variable candidate.

\begin{table}
\bc
\begin{minipage}[]{120mm}
\caption[]{Parameters of the variable candidates in the observed
field.}\end{minipage} \setlength{\tabcolsep}{2.5pt}
\small
 \begin{tabular}{ccccccccccccc}
  \hline\noalign{\smallskip}
ID & RA & DEC & Magnitude NOMAD1&  Amp &  P   & CM$^1$  \\
&(2000) &(2000) & ($B$ ,\, $V$ ,\, $R$) & (mag) & & \\
  \hline\noalign{\smallskip}
C1&04:30:36.780 &+43:43:03.36&14.989, 14.373, 14.450&  0.08 &0.0886       & nm    \\
C2&04:32:58.457 &+43:28:15.92&13.212, 12.620, 12.716&  0.04 &0.0880      & nm    \\
C3&04:33:32.465 &+44:09:09.29&14.551, 13.983, 14.118&  0.14 &5.00        & nm    \\

  \noalign{\smallskip}\hline
\end{tabular}
\ec
\flushleft
  $^1$  Cluster membership: `nm' - non-member.   \\
\label{Table 6}
\end{table}

In this work, the types of variable stars are simply identified according to their light curves appearance. Further multi-color photometry and spectroscopic observations are needed to confirm the exact nature of these stars.

\subsection{Color-magnitude diagrams}

We show the $B-V$ versus $V$ color-magnitude diagram
(CMD) for NGC 1582 in Fig. 9. The variable stars and variable candidates with reliable photometry in all three bands are marked as solid
triangles (red or orange) in this figure. The thick solid lines are the Padova theoretical isochrones (\citealt{Bressan+etal+2012}) with the cluster parameters (t = 300Myr, $V - M_{V}$ = 11.4, $E(B-V)$ = 0.35) from \citet{Baume+etal+2003}. Since the metallicity of NGC 1582 is
unknown, we plotted the isochrones with the given metallicities
of 0.004 (the green line), 0.008 (the blue line) and 0.019 (the
red line).

To discuss whether the variable stars and variable candidates belong to the cluster, we first marked in Fig. 1 the cluster's area with a magenta circle of radius $18.5'$ (this value is suggested by \citealt{Lynga and
Palous+1987}) centered on the cluster center given by \citealt{Dias+etal+2002}. The objects outside the circle are considered as non-members (labeled as $nm$ in Table 3, 4, 5, and 6). V14, a RR Lyrae star, locate right on the edge of the circle. However, it is a field star far behind the cluster (see Section 4.1). In Fig. 9, the variable stars inside the magenta circle of Fig. 1 (V2, V3, V5, V7, V9, V11, V13, V16, V19) are marked with red solid triangles, while the outsiders including V14 marked with orange.

The color-magnitude diagram of NGC 1582 (Fig. 9) is then be used to discuss the membership of the variable stars inside the magenta circle of Fig 1. Stars in the same open cluster have the same age, initial chemical composition, and distance to the Earth (\citealt{Piskunov+etal+2006}; \citealt{Wang+etal+2015}), the stars located in the center dense region of the main sequence within the $3\sigma$ width may therefore be considered as a member star (\citealt{Yang+etal+2013}).

Most of the stars in Fig. 9 are field stars, so the main sequence of NGC 1582 is hard to determine in Fig. 9 because the metallicity of the cluster is still unknown. As shown in Fig. 9, V3, V5, V9 and V11 fit pretty well within the region closed by the three Padova isochrones, which may indicate that they are probable cluster members of NGC 1582. We list them as probable member of the cluster and label as $pm$ in Table 3 and Table 4. Although the remaining stars, V2, V7, V13, V16, and V19 are close to the Padova isochrones with ($Z$) of 0.04 and 0.19, the membership of them is hard to determine. \citet{Yang+etal+2013} argues that V3, V5, V7 (corresponding to V1, V4, and V6 in their paper) are not members of the cluster. However, our observations can not confirm their results. More detailed studies need to be taken in the future.

\begin{figure}[htbp]
   \centering
  \includegraphics[width=0.9\textwidth]{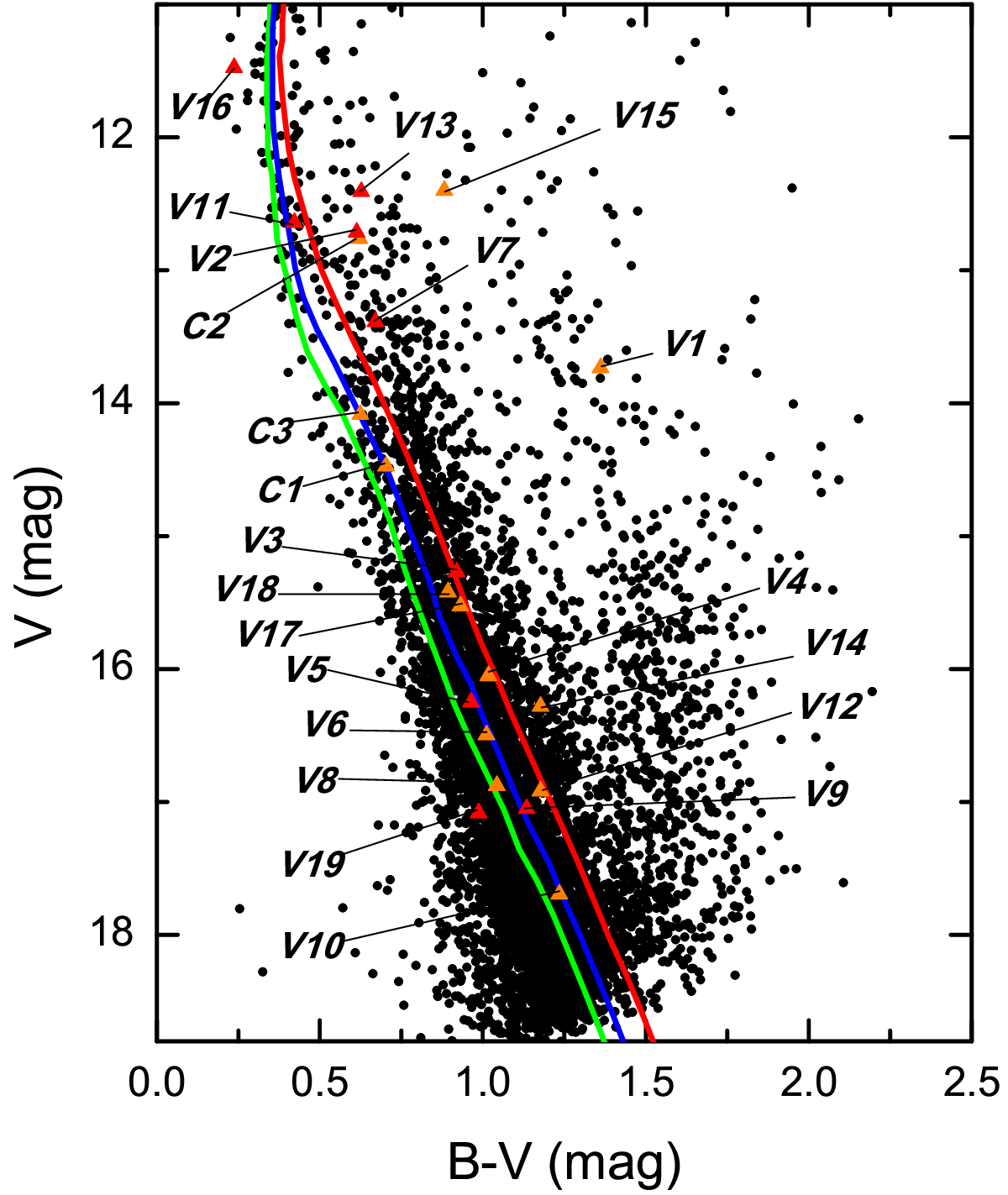}
   \caption{Color-magnitude diagram of NGC 1582. The solid triangles
   indicate the 19 variables stars and 3 variable candidates, in which those inside the magenta circle of Fig.1 with red color, the others with orange. The solid lines represent the Padova theoretical isochrone with the cluster parameters (t = 300 Myr, $V - M_{V}$ = 11.4, $E (B - V) $ = 0.35, the metallicity $Z$ values are 0.004 (the green line), 0.008 (the blue line) and 0.019 (the red line), respectively.).}
   \label{Fig9}
   \end{figure}

\section{Conclusions}

CCD time-series photometry in the $B$, $V$ and $R$ bands by using
Nanshan 1m telescope were performed to study the variable stars in the open
cluster NGC 1582 and its surrounding field. We have detected 22 variable stars in the field of the clusters, out of which 12 are new discoveries and 3 are classified as variable candidates.

Possible types of the variable stars are specified preliminarily using the appearance of their light curves. According to the features on their light curves of the 19 objects, 5 are eclipsing binary systems, and 6 are pulsating variable stars including one certain $\delta$ Scuti star listed in the VSX database and one newly-discovered RR Lyrae star. The type of the others remain unconfirmed, more time-series photometric and spectroscopic observations are necessary.

With a distance of about 8 kpc, the RR Lyrae star (V14) we conclude that it is a field star behind the cluster. 12 objects are distributed over $18.5'$ away from the cluster center, they are considered not the members of the cluster. Based on their locations in the color-magnitude diagram, 4 variable stars are probable members of the cluster. The membership of the remaining 5 variables can not be determined in this work.

 %%%%%%%%%%%%%%%%%%%%%%%%%%%%%%%%%%%%%%%%%%%%%%

\normalem
\begin{acknowledgements}
The authors thank the referee for the very helpful comments. This work is supported by the National Natural Science Foundation of China under Grant No 11273051 and the program of the Light in China
Western Region (LCWR, Grant Nos. XBBS201221 and 2015-XBQN-A-02).
 The CCD photometric data of NGC 1582 were
obtained with the Nanshan 1 m telescope of Xinjiang Astronomical
Observatory. We also acknowledge the use of the archive data from NOMAD1 and GCVS.

\end{acknowledgements}

\bibliographystyle{raa}
\bibliography{bibtex}

\end{document}